\documentclass[11pt]{article}%
\pdfoutput=1
\usepackage{amsmath}
\usepackage{amsfonts}
\usepackage{amssymb}
\usepackage{graphicx}%
\providecommand{\U}[1]{\protect\rule{.1in}{.1in}}
\newtheorem{theorem}{Theorem}

\newtheorem{example}[theorem]{Example}

\begin{document}

\title{Time-dependent one-dimensional electromagnetic wave propagation in inhomogeneous media: exact solution in
terms of transmutations and Neumann series of Bessel functions}
\author{Kira V. Khmelnytskaya$^{1}$, Vladislav V. Kravchenko$^{2}$, Sergii M.
Torba$^{2}$\thanks{The authors acknowledge the support from CONACYT, Mexico
via the projects 222478 and 284470.}\\$^{1}${\footnotesize Faculty of Engineering, Autonomous University of
Queretaro, }\\{\footnotesize Cerro de las Campanas s/n, Quer\'{e}taro,
Qro., CP 76010, M\'{e}xico}\\$^{2}${\footnotesize Department of Mathematics, Cinvestav, Campus
Quer\'{e}taro }\\{\footnotesize Libramiento Norponiente \#2000, Quer\'{e}taro, Qro., CP 76230, M\'{e}xico}\\{\footnotesize khmel@uaq.edu.mx, vkravchenko@math.cinvestav.edu.mx,
storba@math.cinvestav.edu.mx}}
\maketitle

\begin{abstract}
The time-dependent Maxwell system describing electromagnetic wave propagation
in inhomogeneous isotropic media in the one-dimensional case reduces to a
Vekua-type equation for bicomplex-valued functions of a hyperbolic variable,
see \cite{KrRamirez2011}. In \cite{KKT2016JMP} using this reduction a
representation of a general solution of the system was obtained in terms of a
couple of Darboux-associated transmutation operators \cite{KrT2012}. In
\cite{KNT 2015} a Fourier-Legendre expansion of transmutation integral kernels
was obtained. This expansion is used in the present work for obtaining an
exact solution of the problem of the transmission of a normally incident
electromagnetic time-dependent plane wave through an arbitrary inhomogeneous
layer. The result can be used for efficient computation of the transmitted
modulated signals. In particular, it is shown that in the classical situation
of a signal represented in terms of a trigonometric Fourier series the
solution of the problem can be written in the form of Neumann series of Bessel
functions with exact formulas for the coefficients. The representation lends
itself to numerical computation.

\end{abstract}

\section{Introduction}

In the present work, the 1+1 Maxwell system for isotropic inhomogeneous media
is considered. In \cite{KKT2016JMP} a representation for its general solution
was obtained in terms of a couple of Darboux-associated transmutation
(transformation) integral operators (for a recent overview of transmutation
operator theory and applications we refer to \cite{KatrakhovSitnik} and
\cite{SitnikShishkina}). Here we develop further this representation applying
a recent result from \cite{KNT 2015} where a Fourier-Legendre expansion of the
transmutation integral kernels was obtained with explicit formulas for the
coefficients. The main result of the present work is an exact solution of the
classical problem of time-dependent plane electromagnetic wave propagation
through an inhomogeneous medium with an arbitrary profile. The solution has a form
of a series, the coefficients of which are computed recursively. In
particular, if the incoming signal is represented in terms of a Fourier series
the solution of the problem acquires the form of a Neumann series of Bessel
functions. An important feature of this solution representation consists in
the fact that its truncated version approximates the exact solution equally
well for small and for large values of the frequency which means that even
when the partial sums of the Fourier series contain a large number of members
and independently of the largeness of the carrier frequency the approximation
of the solution presented here does not deteriorate.

\section{The 1+1 Maxwell system and the hyperbolic Vekua equation}

The Maxwell system for an isotropic inhomogeneous sourceless medium has the
form%
\begin{align*}
\operatorname{div}(\mu\mathbf{H})&=0,\qquad\operatorname{rot}%
\mathbf{H}=\varepsilon\partial_{t}\mathbf{E},\\
\operatorname{div}(\varepsilon\mathbf{E})&=0,\qquad\operatorname{rot}%
\mathbf{E}=-\mu\partial_{t}\mathbf{H},
\end{align*}
where $\varepsilon$ and $\mu$ are real-valued functions of spatial
coordinates, $\mathbf{E}$ and $\mathbf{H}$ are real-valued vector fields
depending on $t$ and on spatial variables. In the case when all the magnitudes
involved are independent of two spatial coordinates, say, $x_{2}$ and $x_{3}$,
$\varepsilon=\varepsilon(x_{1})$ and $\mu=\operatorname*{Const}$, the
resulting 1+1 Maxwell system for an inhomogeneous medium can be written in the
form
\begin{equation}
\varepsilon(x)\partial_{t}\mathcal{E}=i\partial_{x}\mathcal{H},\quad
i\partial_{x}\mathcal{E}=-\mu\partial_{t}\mathcal{H}\label{Maxwell1+1}%
\end{equation}
where $\mathcal{E}=E_{2}+iE_{3}$, $\mathcal{H}=H_{2}+iH_{3}$, $x=x_{1}$.
Denote $c(x)=1/\sqrt{\varepsilon(x)\mu}$. It is assumed nonvanishing.

In \cite{KrRamirez2011} it was shown that system (\ref{Maxwell1+1}) can be
transformed into the following Vekua-type hyperbolic equation
\begin{equation}
\partial_{\overline{z}}W-\frac{f^{\prime}}{2f}\overline{W}=0 \label{BicompW}%
\end{equation}
where $\partial_{\overline{z}}=\frac{1}{2}(\partial_{\xi}-j\partial_{t})$, $j$
is a hyperbolic imaginary unit, $j^{2}=1$ commuting with $i$, $W$ is a
bicomplex-valued function of the real variables $\xi$ and $t$, $W=u+vj$ and
$u$, $v$ are complex valued (containing the imaginary unit $i$). The function
$f$ is real valued and depends on $\xi$ only. The conjugation with respect to
$j$ is denoted by the bar, $\overline{W}=u-vj$.

The relation between (\ref{Maxwell1+1}) and (\ref{BicompW}) involves the
change of the independent variable $\xi(x)=\sqrt{\mu}\int_{0}^{x}%
\sqrt{\varepsilon(s)}ds$. The function $f$ in (\ref{BicompW}) is related to
$\varepsilon$ and $\mu$ by the equality $f(\xi)=\sqrt{\widetilde{c}(0)}%
/\sqrt{\widetilde{c}(\xi)}$ where and below the tilde means that a function of
$x$ is written as a function of $\xi$, $\widetilde{c}(\xi(x))=c(x)$. The
bicomplex-valued function $W$ is written in terms of $E$ and $H$ as follows%
\begin{equation}
W(\xi,t)=\sqrt{\widetilde{c}(\xi)}\left(  \sqrt{\widetilde{\varepsilon}(\xi
)}\widetilde{\mathcal{E}}(\xi,t)+ij\sqrt{\mu}\widetilde{\mathcal{H}}(\xi,t)\right)  .
\label{Relation W}%
\end{equation}

For the scalar components of a bicomplex-valued function $w=u+vj$ the
following notations will be used
\[
\mathcal{R}(w)=u=\frac{1}{2}(w+\overline{w}),\text{\quad}\mathcal{I}%
(w)=v=\frac{1}{2j}(w-\overline{w})
\]
and
\begin{equation}
w^{\pm}:=\mathcal{R}\left(  w\right)  \pm\mathcal{I}(w). \label{W+-}%
\end{equation}

Then from (\ref{Relation W}) we have
\begin{equation}
\widetilde{\mathcal{E}}(\xi,t)=\frac{1}{\sqrt{\widetilde{c}(\xi)\widetilde
{\varepsilon}(\xi)}}\mathcal{R}(W(\xi,t)) \label{E tilde}%
\end{equation}
and
\begin{equation}
\widetilde{\mathcal{H}}(\xi,t)=-\frac{i}{\sqrt{\widetilde{c}(\xi)\mu}%
}\mathcal{I}(W(\xi,t)). \label{H tilde}%
\end{equation}

\section{A general solution of the hyperbolic Vekua equation}

Together with the Vekua equation (\ref{BicompW}) consider its special case,
the hyperbolic Cauchy-Riemann system
\begin{equation}
\partial_{\overline{z}}w=0 \label{CR hyper}%
\end{equation}
which was studied in several publications (see, e.g., \cite{Lavrentyev and
Shabat}, \cite{MotterRosa}, \cite{Wen} and more recent \cite{KrT-CAOT}). Its
general solution can be written in the form%
\[
w(\xi,t)=P^{+}\Phi(t+\xi)+P^{-}\Psi(t-\xi)
\]
where $\Phi$ and $\Psi$ are arbitrary continuously differentiable scalar
functions (complex valued containing the imaginary unit $i$) and $P^{\pm
}=\frac{1}{2}\left(  1\pm j\right)  $.

In \cite{KrT-CAOT} there was established a relation between solutions of
(\ref{CR hyper}) and (\ref{BicompW}). Any solution of (\ref{BicompW}) can be
represented in the form
\begin{equation}
W(\xi,t)=T_{f}\left[  \mathcal{R}\left(  w(\xi,t)\right)  \right]
+jT_{1/f}\left[  \mathcal{I}\left(  w(\xi,t)\right)  \right]
\label{W transmut w}%
\end{equation}
where $w$ is a solution of (\ref{CR hyper}), $T_{f}$ and $T_{1/f}$ are
Darboux-associated transmutation operators defined in \cite{KrT2012}, see also
\cite{CKM2012} and \cite{KrT-CAOT}. The operators $T_{f}$ and $T_{1/f}$ are
applied with respect to the variable $\xi$ and have the form of second-kind
Volterra integral operators,
\[
T_{f}u(\xi)=u(\xi)+\int_{-\xi}^{\xi}\mathbf{K}_{f}(\xi,\tau)u(\tau)d\tau
\]
and
\[
T_{1/f}u(\xi)=u(\xi)+\int_{-\xi}^{\xi}\mathbf{K}_{1/f}(\xi,\tau)u(\tau)d\tau
\]
with continuously differentiable kernels $\mathbf{K}_{f}$ and $\mathbf{K}%
_{1/f}$.

In \cite{KNT 2015} a representation of the transmutation kernels in the form
of Fourier-Legendre series was obtained. Namely, the transmutation kernels
$\mathbf{K}_{f}$ and $\mathbf{K}_{1/f}$ have the form
\begin{equation}
\mathbf{K}_{f}(\xi,\tau)=\sum_{n=0}^{\infty}\frac{a_{n}(\xi)}{\xi}P_{n}\left(
\frac{\tau}{\xi}\right)  \label{Kf}%
\end{equation}
and
\begin{equation}
\mathbf{K}_{1/f}(\xi,\tau)=\sum_{n=0}^{\infty}\frac{b_{n}(\xi)}{\xi}%
P_{n}\left(  \frac{\tau}{\xi}\right)  \label{K1/f}%
\end{equation}
where $P_{n}$ stands for the Legendre polynomial of order $n$, for every
$\xi>0$ the series converge uniformly with respect to $\tau\in\left[  -\xi
,\xi\right]  $, and for the coefficients $a_{n}$ and $b_{n}$, $n=0,1,2,\ldots$
explicit formulas are obtained. In order to write them down we introduce the
systems of functions $\left\{  \varphi_{k}\right\}  _{k=0}^{\infty}$ and
$\left\{  \psi_{k}\right\}  _{k=0}^{\infty}$ defined as follows.

Consider two sequences of recursive integrals%
\begin{equation}
X^{(0)}\equiv1,\qquad X^{(n)}(\xi)=n\int_{0}^{\xi}X^{(n-1)}(s)\left(
f^{2}(s)\right)  ^{(-1)^{n}}\,\mathrm{d}s,\qquad n=1,2,\ldots\label{Xn}%
\end{equation}
and
\begin{equation}
\widetilde{X}^{(0)}\equiv1,\qquad\widetilde{X}^{(n)}(\xi)=n\int_{0}^{\xi
}\widetilde{X}^{(n-1)}(s)\left(  f^{2}(s)\right)  ^{(-1)^{n-1}}\,\mathrm{d}%
s,\qquad n=1,2,\ldots. \label{Xtiln}%
\end{equation}

The two families of functions $\left\{  \varphi_{k}\right\}  _{k=0}^{\infty}$
and $\left\{  \psi_{k}\right\}  _{k=0}^{\infty}$ are constructed according to
the rules
\begin{equation}
\varphi_{k}(\xi)=%
\begin{cases}
f(\xi)X^{(k)}(\xi), & k\text{\ odd},\\
f(\xi)\widetilde{X}^{(k)}(\xi), & k\text{\ even},
\end{cases}
\label{phik}%
\end{equation}
and
\begin{equation}
\psi_{k}(\xi)=%
\begin{cases}
\dfrac{\widetilde{X}^{(k)}(\xi)}{f(\xi)}, & k\text{\ odd,}\\
\dfrac{X^{(k)}(\xi)}{f(\xi)}, & k\text{\ even}.
\end{cases}
\label{psik}%
\end{equation}

The coefficients $a_{n}$ and $b_{n}$ in (\ref{Kf}) and (\ref{K1/f}) admit the
following representation \cite{KNT 2015}%
\[
a_{n}(\xi)=\frac{2n+1}{2}\left(  \sum_{k=0}^{n}\frac{l_{k,n}\varphi_{k}(\xi
)}{\xi^{k}}-1\right)
\]
and
\[
b_{n}(\xi)=\frac{2n+1}{2}\left(  \sum_{k=0}^{n}\frac{l_{k,n}\psi_{k}(\xi)}%
{\xi^{k}}-1\right)
\]
where $l_{k,n}$ is $k$-th power's coefficient of the Legendre polynomial of
order $n$, that is $P_{n}(x)=\sum_{k=0}^{n}l_{k,n}x^{k}$.

Besides these direct formulas for the coefficients $a_{n}$ and $b_{n}$, in
\cite{KNT 2015} a recurrent integration procedure for their computation was
proposed, convenient for numerical applications.

Substitution of (\ref{Kf}) and (\ref{K1/f}) into (\ref{W transmut w}) leads to
the following representation of a general solution of (\ref{BicompW})%
\[
W(\xi,t)=w(\xi,t)+\sum_{n=0}^{\infty}\frac{a_{n}(\xi)}{\xi}\int_{-\xi}^{\xi
}P_{n}\left(  \frac{\tau}{\xi}\right)  \mathcal{R}\left(  w(\tau,t)\right)
\,\mathrm{d}\tau+j\sum_{n=0}^{\infty}\frac{b_{n}(\xi)}{\xi}\int_{-\xi}^{\xi
}P_{n}\left(  \frac{\tau}{\xi}\right)  \mathcal{I}\left(  w(\tau,t)\right)
\,\mathrm{d}\tau
\]
where $w$ is a general solution of (\ref{CR hyper}).

\section{Normally incident plane wave propagation through an inhomogeneous
medium}

In this section we study the classical problem of a normally incident plane
wave propagation through an inhomogeneous medium (see, e.g., \cite[Chapter
8]{OstrovskyPotapov}). The electromagnetic field $\mathcal{E}$ and
$\mathcal{H}$ satisfying (\ref{Maxwell1+1}) is supposed to be known at $x=0$,%
\begin{equation}
\mathcal{E}(0,t)=\mathcal{E}_{0}(t)\quad\text{and}\quad\mathcal{H}%
(0,t)=\mathcal{H}_{0}(t),\qquad t\in\lbrack\alpha,\beta].
\label{init cond}%
\end{equation}
We assume $\mathcal{E}_{0}$ and $\mathcal{H}_{0}$ to be continuously
differentiable functions.

Problem (\ref{Maxwell1+1}), (\ref{init cond}) can be reformulated in terms of
the function (\ref{Relation W}). Find a solution of (\ref{BicompW}) satisfying
the condition%
\begin{equation}
W(0,t)=W_{0}(t)\label{initial condition}%
\end{equation}
where
\begin{equation}
W_{0}=\sqrt{c(0)\varepsilon(0)}\mathcal{E}_{0}+ij\sqrt{c(0)\mu}\mathcal{H}%
_{0}\label{W0}%
\end{equation}
is a given continuously differentiable function. Then the following statement
is valid.

\begin{theorem} The unique solution of Problem (\ref{BicompW}),
(\ref{initial condition}) has the form
\begin{equation}\label{complete sol}
\begin{split}
W(\xi,t) &  =\frac{1}{2}\left(  W_{0}^{+}(t+\xi)+W_{0}^{-}(t-\xi)\right)
+\frac{1}{2}\sum_{n=0}^{\infty}\frac{a_{n}(\xi)}{\xi}\int_{-\xi}^{\xi}%
P_{n}\left(  \frac{\tau}{\xi}\right)  \left(  W_{0}^{+}(t+\tau)+W_{0}%
^{-}(t-\tau)\right)  \,\mathrm{d}\tau\\
&\quad  +\frac{j}{2}\left(  W_{0}^{+}(t+\xi)-W_{0}^{-}(t-\xi)\right)  +\frac{j}%
{2}\sum_{n=0}^{\infty}\frac{b_{n}(\xi)}{\xi}\int_{-\xi}^{\xi}P_{n}\left(
\frac{\tau}{\xi}\right)  \left(  W_{0}^{+}(t+\tau)-W_{0}^{-}(t-\tau)\right)
\,\mathrm{d}\tau\\
&  =P^{+}W_{0}^{+}(t+\xi)+P^{-}W_{0}^{-}(t-\xi)+\frac{1}{2\xi}\sum
_{n=0}^{\infty}( a_{n}(\xi)+jb_{n}(\xi))  \int_{-\xi}^{\xi}%
P_{n}\left(  \frac{\tau}{\xi}\right)  W_{0}^{+}(t+\tau)\,\mathrm{d}%
\tau\\
&\quad  +\frac{1}{2\xi}\sum_{n=0}^{\infty}(  a_{n}(\xi)-jb_{n}(\xi))
\int_{-\xi}^{\xi}P_{n}\left(  \frac{\tau}{\xi}\right)  W_{0}^{-}%
(t-\tau)\,\mathrm{d}\tau,
\end{split}
\end{equation}
from which the unique solution of Problem (\ref{Maxwell1+1}), (\ref{init cond}%
) is obtained by means of (\ref{E tilde}), (\ref{H tilde}) and an inverse
change of the variable $\xi\rightarrow x$.
\end{theorem}

\noindent\textbf{Proof. }The unique solution of Problem (\ref{BicompW}),
(\ref{initial condition}) has the form \cite{KKT2016JMP}
\begin{equation}
W(\xi,t)=\frac{1}{2}T_{f}\left[  W_{0}^{+}(t+\xi)+W_{0}^{-}(t-\xi)\right]
+\frac{j}{2}T_{1/f}\left[  W_{0}^{+}(t+\xi)-W_{0}^{-}(t-\xi)\right]
.\label{Solution initial problem}%
\end{equation}
Due to (\ref{Kf}) and (\ref{K1/f}) this can be written in the form
(\ref{complete sol}). $\blacksquare$

For some practically interesting initial data the integrals in
(\ref{complete sol}) can be evaluated in a closed form. Let us consider one
such example corresponding to modulated electromagnetic waves which are
represented as partial sums of Fourier series
\begin{equation}
\mathcal{E}_{0}(t)=\sum_{m=-M}^{M}\alpha_{m}e^{i(\omega_{0}+m\omega)t}%
\qquad\text{and}\qquad\mathcal{H}_{0}(t)=\sum_{m=-M}^{M}\beta_{m}e^{i(\omega
_{0}+m\omega)t}. \label{EHmodulated}%
\end{equation}
This leads to a similar form for the initial data $W_{0}$ in
(\ref{initial condition}),
\begin{equation}
W_{0}(t)=\sum_{m=-M}^{M}c_{m}e^{i(\omega_{0}+m\omega)t} \label{W0exp}%
\end{equation}
where the bicomplex numbers $c_{m}$ are related to $\alpha_{m}$, $\beta_{m}%
\in\mathbb{C}$ as follows%
\[
c_{m}=\sqrt{c(0)}\left(  \sqrt{\varepsilon(0)}\alpha_{m}+ij\sqrt{\mu}\beta
_{m}\right)  .
\]

Note that $W_{0}^{\pm}(t\pm\xi)=\sum_{m=-M}^{M}c_{m}^{\pm}e^{i(\omega
_{0}+m\omega)(t\pm\xi)}$ and due to (\ref{complete sol}), the solution of
Problem (\ref{BicompW}), (\ref{initial condition}) with the initial data
(\ref{W0exp}) is given by the formula%
\begin{equation}\label{W sol1}
\begin{split}
W(\xi,t)  &  =P^{+}\sum_{m=-M}^{M}c_{m}^{+}e^{i(\omega_{0}+m\omega)(t+\xi
)}+P^{-}\sum_{m=-M}^{M}c_{m}^{-}e^{i(\omega_{0}+m\omega)(t-\xi)}\\
&\quad  +\frac{1}{2\xi}\sum_{n=0}^{\infty}\left(  a_{n}(\xi)+jb_{n}(\xi)\right)
\int_{-\xi}^{\xi}P_{n}\left(  \frac{\tau}{\xi}\right)  \sum_{m=-M}^{M}%
c_{m}^{+}e^{i(\omega_{0}+m\omega)\left(  t+\tau\right)  }\,\mathrm{d}%
\tau\\
&\quad  +\frac{1}{2\xi}\sum_{n=0}^{\infty}\left(  a_{n}(\xi)-jb_{n}(\xi)\right)
\int_{-\xi}^{\xi}P_{n}\left(  \frac{\tau}{\xi}\right)  \sum_{m=-M}^{M}%
c_{m}^{-}e^{i(\omega_{0}+m\omega)\left(  t-\tau\right)  }\,\mathrm{d}\tau.
\end{split}
\end{equation}
Using formula 2.17.5(1) from \cite{Prudnikov} the integrals here can be
evaluated,%
\[
\int_{-\xi}^{\xi}P_{n}\left(  \frac{\tau}{\xi}\right)  e^{i(\omega_{0}%
+m\omega)\tau}\,\mathrm{d}\tau=2\xi e^{\frac{n\pi i}{2}}j_{n}\left(
(\omega_{0}+m\omega)\xi\right)
\]
and
\[
\int_{-\xi}^{\xi}P_{n}\left(  \frac{\tau}{\xi}\right)  e^{-i(\omega
_{0}+m\omega)\tau}\,\mathrm{d}\tau=\left(  -1\right)  ^{n}2\xi e^{\frac{n\pi
i}{2}}j_{n}\left(  (\omega_{0}+m\omega)\xi\right)
\]
where $j_{n}$ stands for the spherical Bessel function of order $n$. Thus, the
bicomplex-valued function $W$ from (\ref{W sol1}) takes the form%
\begin{equation*}
\begin{split}
W(\xi,t)  &  =\sum_{m=-M}^{M}e^{i(\omega_{0}+m\omega)t}\biggl(  P^{+}c_{m}%
^{+}e^{i(\omega_{0}+m\omega)\xi}+P^{-}c_{m}^{-}e^{-i(\omega_{0}+m\omega)\xi
} \\
&\quad  +c_{m}^{+}\sum_{n=0}^{\infty}\left(  a_{n}(\xi)+jb_{n}(\xi)\right)
e^{\frac{n\pi i}{2}}j_{n}\left(  (\omega_{0}+m\omega)\xi\right) \\
& \quad   +c_{m}^{-}\sum_{n=0}^{\infty}\left(  -1\right)  ^{n}\left(
a_{n}(\xi)-jb_{n}(\xi)\right)  e^{\frac{n\pi i}{2}}j_{n}\left(  (\omega
_{0}+m\omega)\xi\right)  \biggr)  .
\end{split}
\end{equation*}

In order to obtain the corresponding formula for the solution of Problem
(\ref{Maxwell1+1}), (\ref{init cond}) with the initial data of the form
(\ref{EHmodulated}), we notice that
\begin{equation*}
\begin{split}
\mathcal{R}\left(  W(\xi,t)\right)   &  =\sum_{m=-M}^{M}e^{i(\omega
_{0}+m\omega)t}\biggl(  \frac{c_{m}^{+}}{2}e^{i(\omega_{0}+m\omega)\xi}%
+\frac{c_{m}^{-}}{2}e^{-i(\omega_{0}+m\omega)\xi}  \\
&\quad  +c_{m}^{+}\sum_{n=0}^{\infty}a_{n}(\xi)e^{\frac{n\pi i}{2}}%
j_{n}\left(  (\omega_{0}+m\omega)\xi\right)  +c_{m}^{-}\sum_{n=0}^{\infty
}\left(  -1\right)  ^{n}a_{n}(\xi)e^{\frac{n\pi i}{2}}j_{n}\left(  (\omega
_{0}+m\omega)\xi\right)  \biggl)
\end{split}
\end{equation*}
and
\begin{equation*}
\begin{split}
\mathcal{I}\left(  W(\xi,t)\right)   &  =\sum_{m=-M}^{M}e^{i(\omega
_{0}+m\omega)t}\biggl(  \frac{c_{m}^{+}}{2}e^{i(\omega_{0}+m\omega)\xi}%
-\frac{c_{m}^{-}}{2}e^{-i(\omega_{0}+m\omega)\xi}  \\
&\quad   +c_{m}^{+}\sum_{n=0}^{\infty}b_{n}(\xi)e^{\frac{n\pi i}{2}}%
j_{n}\left(  (\omega_{0}+m\omega)\xi\right)  -c_{m}^{-}\sum_{n=0}^{\infty
}\left(  -1\right)  ^{n}b_{n}(\xi)e^{\frac{n\pi i}{2}}j_{n}\left(  (\omega
_{0}+m\omega)\xi\right)  \biggr)  .
\end{split}
\end{equation*}
Hence due to (\ref{E tilde}) and (\ref{H tilde}) we obtain the following result.

\begin{theorem} The unique solution of Problem (\ref{Maxwell1+1}),
(\ref{init cond}) with the initial data of the form (\ref{EHmodulated})
written in terms of coordinates $\xi$ and $t$ has the form%
\begin{equation}\label{E tilde modulated}%
\begin{split}
\widetilde{\mathcal{E}}(\xi,t) &  =\frac{1}{\sqrt{\widetilde{c}(\xi
)\widetilde{\varepsilon}(\xi)}}\sum_{m=-M}^{M}e^{i(\omega_{0}+m\omega
)t}\biggl(  \frac{c_{m}^{+}}{2}e^{i(\omega_{0}+m\omega)\xi}+\frac{c_{m}^{-}}%
{2}e^{-i(\omega_{0}+m\omega)\xi}\\
&  \quad  +c_{m}^{+}\sum_{n=0}^{\infty}a_{n}(\xi)e^{\frac{n\pi i}{2}}%
j_{n}\left(  (\omega_{0}+m\omega)\xi\right)  +c_{m}^{-}\sum_{n=0}^{\infty
}\left(  -1\right)  ^{n}a_{n}(\xi)e^{\frac{n\pi i}{2}}j_{n}\left(  (\omega
_{0}+m\omega)\xi\right)  \biggr)
\end{split}
\end{equation}
and
\begin{equation}\label{H tilde modulated}%
\begin{split}
\widetilde{\mathcal{H}}(\xi,t) &  =-\frac{i}{\sqrt{\widetilde{c}(\xi)\mu}}%
\sum_{m=-M}^{M}e^{i(\omega_{0}+m\omega)t}\biggl(  \frac{c_{m}^{+}}%
{2}e^{i(\omega_{0}+m\omega)\xi}-\frac{c_{m}^{-}}{2}e^{-i(\omega_{0}%
+m\omega)\xi}\\
&  \quad  +c_{m}^{+}\sum_{n=0}^{\infty}b_{n}(\xi)e^{\frac{n\pi i}{2}}%
j_{n}\left(  (\omega_{0}+m\omega)\xi\right)  -c_{m}^{-}\sum_{n=0}^{\infty
}\left(  -1\right)  ^{n}b_{n}(\xi)e^{\frac{n\pi i}{2}}j_{n}\left(  (\omega
_{0}+m\omega)\xi\right)  \biggr)  .
\end{split}
\end{equation}
\end{theorem}

Thus, the solution of Problem (\ref{Maxwell1+1}), (\ref{init cond}) with the
initial data of the form (\ref{EHmodulated}) is obtained in the form of a
Neumann series of Bessel functions (see, e.g., \cite{Watson}, \cite{Wilkins}
and a recent publication on the subject \cite{Baricz et al} and references
therein). Let us discuss an important feature of this solution representation.
Namely, for practical computation one needs to consider the truncated series
in (\ref{E tilde modulated}) and (\ref{H tilde modulated}). Thus, together
with the exact solution (\ref{E tilde modulated}), (\ref{H tilde modulated})
consider its approximation%
\begin{equation}\label{E tildeN}%
\begin{split}
\widetilde{\mathcal{E}}_{N}(\xi,t) &  :=\frac{1}{\sqrt{\widetilde{c}%
(\xi)\widetilde{\varepsilon}(\xi)}}\sum_{m=-M}^{M}e^{i(\omega_{0}+m\omega
)t}\biggl(  \frac{c_{m}^{+}}{2}e^{i(\omega_{0}+m\omega)\xi}+\frac{c_{m}^{-}}%
{2}e^{-i(\omega_{0}+m\omega)\xi}\\
&  \quad  +c_{m}^{+}\sum_{n=0}^{N}a_{n}(\xi)e^{\frac{n\pi i}{2}}j_{n}\left(
(\omega_{0}+m\omega)\xi\right)  +c_{m}^{-}\sum_{n=0}^{N}\left(  -1\right)
^{n}a_{n}(\xi)e^{\frac{n\pi i}{2}}j_{n}\left(  (\omega_{0}+m\omega)\xi\right)
\biggr)
\end{split}
\end{equation}
and%
\begin{equation}\label{H tildeN}%
\begin{split}
\widetilde{\mathcal{H}}_{N}(\xi,t) &  =-\frac{i}{\sqrt{\widetilde{c}(\xi)\mu}%
}\sum_{m=-M}^{M}e^{i(\omega_{0}+m\omega)t}\biggl(  \frac{c_{m}^{+}}%
{2}e^{i(\omega_{0}+m\omega)\xi}-\frac{c_{m}^{-}}{2}e^{-i(\omega_{0}%
+m\omega)\xi}\\
&  \quad  +c_{m}^{+}\sum_{n=0}^{N}b_{n}(\xi)e^{\frac{n\pi i}{2}}j_{n}\left(
(\omega_{0}+m\omega)\xi\right)  -c_{m}^{-}\sum_{n=0}^{N}\left(  -1\right)
^{n}b_{n}(\xi)e^{\frac{n\pi i}{2}}j_{n}\left(  (\omega_{0}+m\omega)\xi\right)
\biggr)  .
\end{split}
\end{equation}
It is important to note that for real $\omega_{0}$ and $\omega$ the difference
between the exact solution and the approximate one is independent of the
largeness of $\omega_{0}$ and $\omega$. More precisely, consider
\[
\left\vert \widetilde{\mathcal{E}}(\xi,t)-\widetilde{\mathcal{E}}_{N}%
(\xi,t)\right\vert =\left\vert \frac{1}{\sqrt{\widetilde{c}(\xi)\widetilde
{\varepsilon}(\xi)}}\sum_{m=-M}^{M}e^{i(\omega_{0}+m\omega)t}\left(  c_{m}%
^{+}A_{1;N}(\omega,\xi)+c_{m}^{-}A_{2;N}(\omega,\xi)\right)  \right\vert
\]
where $A_{1;N}(\omega,\xi):=\sum_{n=N+1}^{\infty}a_{n}(\xi)e^{\frac{n\pi i}{2}}j_{n}\left(  (\omega_{0}+m\omega)\xi\right)  $ and $A_{2;N}(\omega
,\xi):=\sum_{n=N+1}^{\infty}\left(  -1\right)  ^{n}a_{n}(\xi)e^{\frac{n\pi
i}{2}}$ $\times j_{n}\left(  (\omega_{0}+m\omega)\xi\right)  $. Due to \cite[Theorem
4.1]{KNT 2015} one has $\left\vert A_{1,2;N}(\omega,\xi)\right\vert
\leq\varepsilon_{N}(\xi)$ where $\varepsilon_{N}$ is a positive function which
can be made arbitrarily small choosing a sufficiently large $N$. Thus,
\begin{equation}
\left\vert \widetilde{\mathcal{E}}(\xi,t)-\widetilde{\mathcal{E}}_{N}%
(\xi,t)\right\vert \leq\frac{\varepsilon_{N}(\xi)}{\sqrt{\widetilde{c}%
(\xi)\widetilde{\varepsilon}(\xi)}}\sum_{m=-M}^{M}\left(  \left\vert c_{m}%
^{+}\right\vert +\left\vert c_{m}^{-}\right\vert \right)  .\label{Estimate E}%
\end{equation}
A similar reasoning is applicable to (\ref{H tilde modulated}) and
(\ref{H tildeN}). One obtains an estimate of the form
\begin{equation}
\left\vert \widetilde{\mathcal{H}}(\xi,t)-\widetilde{\mathcal{H}}_{N}%
(\xi,t)\right\vert \leq\frac{\varepsilon_{N}(\xi)}{\sqrt{\widetilde{c}(\xi
)\mu}}\sum_{m=-M}^{M}\left(  \left\vert c_{m}^{+}\right\vert +\left\vert
c_{m}^{-}\right\vert \right)  .\label{Estimate H}%
\end{equation}
The estimates (\ref{Estimate E}) and (\ref{Estimate H}) imply that the
approximations (\ref{E tildeN}) and (\ref{H tildeN}) perform equally well for
small and for large values of $\omega_{0}$ and $\left\vert m\omega\right\vert
$.

\begin{example}
As an illustration let us consider a situation when the transmutation kernels
$\mathbf{K}_{f}$ and $\mathbf{K}_{1/f}$ admit finite sum
representations of the form (\ref{Kf}) and (\ref{K1/f}) and hence the solution \eqref{E tilde modulated}, \eqref{H tilde modulated} admits a closed form expression. As was observed in
\cite{KrT2012}, when
\begin{equation}
f(\xi)=\frac{1}{(1+\xi)^{2}} \label{f example}%
\end{equation}
the integral kernels of the transmutation operators $T_{f}$ and $T_{1/f}$ are
given by
\[
\mathbf{K}_{f}(\xi,t)=\frac{(3t-1)(\xi+1)^{2}-3(t-1)^{2}(t+1)}{4(\xi+1)^{2}%
}\quad\text{and}\quad\mathbf{K}_{1/f}(\xi,t)=\frac{3\xi^{2}+6\xi+4-3t^{2}%
+2t}{4(\xi+1)}.
\]
They are polynomials with respect to the variable $t$ and hence admit a
representation in the form of a finite linear combination of the Legendre
polynomials. Simple calculation gives us the following formulas for the
coefficients $\left\{  a_{n}\right\}  $ and $\left\{  b_{n}\right\}  $ in
(\ref{Kf}) and (\ref{K1/f}). We have
\[
a_{0}(\xi)=-\frac{\xi\left(  \xi+2\right)  }{2\left(  \xi+1\right)  ^{2}%
},\quad a_{1}(\xi)=\frac{3\xi^{2}\left(  \xi^{2}+5\xi+5\right)  }{10\left(
\xi+1\right)  ^{2}},\quad a_{2}(\xi)=\frac{\xi^{3}}{2\left(  \xi+1\right)
^{2}},\quad a_{3}(\xi)=-\frac{3\xi^{4}}{10\left(  \xi+1\right)  ^{2}},
\]
$\quad a_{n}\equiv0,\,n=4,5,\ldots$ and
\[
b_{0}(\xi)=\frac{\xi\left(  \xi+2\right)  }{2},\quad b_{1}(\xi)=\frac{\xi^{2}%
}{2\left(  \xi+1\right)  },\quad b_{2}(\xi)=-\frac{\xi^{3}}{2\left(
\xi+1\right)  },\quad b_{n}\equiv0,\,n=3,4,\ldots.
\]
As was shown in \cite{KKT2016JMP} this example is associated with the
following electromagnetic system. Consider system (\ref{Maxwell1+1}) with the
permittivity of the form $\varepsilon(x)=(\alpha x+\beta)^{-2+2\ell}$ where
$\ell\neq0$ and $\alpha$, $\beta\in\mathbb{R}$ are such that $\alpha
x+\beta>0$ on the interval of interest. Then $\xi(x)=\sqrt{\mu}\int_{0}%
^{x}\sqrt{\varepsilon(s)}\,ds=\frac{\sqrt{\mu}}{\alpha\ell}\left(  (\alpha
x+\beta)^{\ell}-\beta^{\ell}\right)  $. Hence
\[
x=\frac{\beta}{\alpha}\left(  \left(  1+\frac{\alpha\ell\xi}{\sqrt{\mu}%
\beta^{\ell}}\right)  ^{1/\ell}-1\right)  ,\quad\widetilde{c}(\xi)=\frac
{1}{\sqrt{\mu}}\left(  \frac{\alpha\ell\xi}{\sqrt{\mu}}+\beta^{\ell}\right)
^{(1-\ell)/\ell}\quad\text{and}\quad f(\xi)=\frac{1}{\bigl(1+\frac{\alpha\ell
}{\sqrt{\mu}\beta^{\ell}}\xi\bigr)^{\frac{1-\ell}{2\ell}}}.
\]
Thus, take $\ell=1/5$, $\alpha=5$, $\beta=1$ and $\mu=1$. Then $f$ has the
form (\ref{f example}), and solution (\ref{E tilde modulated}) and
(\ref{H tilde modulated}) of Problem (\ref{Maxwell1+1}), (\ref{init cond})
with the initial data of the form (\ref{EHmodulated}) admits a finite sum
representation obtained from (\ref{E tilde modulated}) and
(\ref{H tilde modulated}) by substituting the coefficients $\left\{
a_{n}\right\}  $ and $\left\{  b_{n}\right\}  $ from this example.
\end{example}

When neither the integrals in \eqref{complete sol} can be evaluated in the closed form, nor initial data $\mathcal{E}_0$ and $\mathcal{H}_0$ can be sufficiently closely approximated by partial sums of Fourier series \eqref{EHmodulated}, one faces the problem of efficient numerical evaluation of the integrals in \eqref{complete sol} for all required pairs of $t$ and $\xi$. One of the possibilities is to proceed as follows. Denote by $W_N(\xi,t)$ the truncated expression \eqref{complete sol}. We use the explicit representation of the Legendre polynomial, $P_n(x)=\sum_{k=0}^n l_{k,n}x^k$ to obtain that
\begin{equation*}
    \begin{split}
       W_N(\xi,t) & = P^+W_0^+(t+\xi) + P^-W_0^-(t-\xi) + \frac 1{2\xi}\sum_{n=0}^N\bigl(a_n(\xi)+jb_n(\xi)\bigr) \sum_{k=0}^n\frac{l_{k,n}}{\xi^k}\int_{-\xi}^\xi \tau^k W_0^+(t+\tau)\,\mathrm{d}\tau \\
         & \quad + \frac 1{2\xi}\sum_{n=0}^N\bigl(a_n(\xi)-jb_n(\xi)\bigr) \sum_{k=0}^n\frac{l_{k,n}}{\xi^k}\int_{-\xi}^\xi \tau^k W_0^-(t-\tau)\,\mathrm{d}\tau.
     \end{split}
\end{equation*}
Now we proceed analogously to the derivation of formula (5.2) from \cite{KKT2016JMP},
\[
\int_{-\xi}^\xi \tau^k W_0^+(t+\tau)\,\mathrm{d}\tau = \sum_{l=0}^k \binom{k}{l} (-1)^{k-l} t^{k-l}\int_{t-\xi}^{t+\xi} z^l W_0^+(z)\,\mathrm{d}z,
\]
and similarly for the second integral. Hence
\begin{equation}\label{W_Nrearranged}
\begin{split}
    W_N(\xi,t) &= P^+W_0^+(t+\xi)  +\sum_{l=0}^N\Biggl(\int_{t-\xi}^{t+\xi}z^l W_0^+(z)\,\mathrm{d}z\Biggr) \sum_{k=l}^N c_{k,N}^+(\xi)\binom{k}{l} (-t)^{k-l}\\
           &\quad + P^-W_0^-(t-\xi)+\sum_{l=0}^N(-1)^l \Biggl(\int_{t-\xi}^{t+\xi}z^l W_0^-(z)\,\mathrm{d}z\Biggr) \sum_{k=l}^N c_{k,N}^-(\xi)\binom{k}{l} t^{k-l},
\end{split}
\end{equation}
where
\[
c_{k,N}^\pm(\xi) = \frac{1}{\xi^{k+1}}\sum_{n=k}^N\frac{l_{k,n}}2\bigl(a_n(\xi)\pm j b_n(\xi)\bigr).
\]

By such rearrangement we replace the problem of evaluation of the integrals in \eqref{complete sol}, where for each pair of $t$ and $\xi$ one needs to integrate different functions $P_n\bigl(\frac{\cdot}{\xi}\bigr)W_0^\pm(t\pm\cdot)$, by the problem of evaluation of the integrals of the same functions $z^kW_0^\pm(z)$ over different segments. Taking into account that the indefinite integrals $\int_0^x z^k W_0^\pm(z)\,\mathrm{d}z$, $k=0,\ldots,N$, can be efficiently computed numerically via, e.g., piece-wise polynomial interpolation of the integrands (see, e.g.,  \cite[Section 2.13]{Rabinowitz}), formula \eqref{W_Nrearranged} may significantly reduce the computational cost. However, one may expect the loss of precision due to multiplication by large coefficients $l_{k,n} \binom{k}{l}$ and near $\xi=0$ due to division by $\xi^{k+1}$.

\begin{example}\label{Ex1}
For the numerical test we consider the problem from Example 6.2 \cite{KKT2016JMP}. It consists in the system (\ref{Maxwell1+1}) with the
permittivity of the form
\begin{equation}\label{ExEpsilon}
\varepsilon(x)=(\alpha x+\beta)^{-2},
\end{equation}
where $\alpha$
and $\beta$ are some real numbers, such that $\alpha x+\beta$ does not vanish
on the interval of interest. Then $\xi=\sqrt{\mu}\int
_{0}^{x}\sqrt{\varepsilon(s)}ds=\frac{\sqrt{\mu}}{\alpha}\log\frac{\alpha
x+\beta}{\beta}$. Hence
\[
x=\frac{\beta}{\alpha}\left(  e^{\frac{\alpha\xi}{\sqrt{\mu}}}-1\right)
\quad\text{and}\quad\widetilde{\varepsilon}(\xi)=\frac{1}{\beta^{2}}%
e^{-\frac{2\alpha\xi}{\sqrt{\mu}}},\quad\widetilde{c}(\xi)=\frac{\beta}%
{\sqrt{\mu}}e^{\frac{\alpha\xi}{\sqrt{\mu}}},\quad f(\xi)=e^{-\frac{\alpha\xi
}{2\sqrt{\mu}}}.
\]
In this case the Vekua equation (\ref{BicompW}) has the form
\begin{equation}
\partial_{\overline{z}}W+\gamma\overline{W}=0 \label{VekuaConst}%
\end{equation}
where the coefficient $\gamma$ is constant, $\gamma=\alpha/\left(  4\sqrt{\mu
}\right)  $.

The Vekua equation \eqref{VekuaConst} possesses an exact solution of the form \cite{KKT2016JMP}
\begin{equation}\label{Ex2Wxit}
W(\xi,t)=Ae^{i\Omega t}\left(  e^{D\xi}+\frac{D+C}{D-C}e^{-D\xi}%
+\frac{2ij\Omega}{D-C}\sinh D\xi\right)  ,
\end{equation}
here $A$ and $\Omega$ are arbitrary constants, $C=\frac{\alpha}{2\sqrt{\mu}}$,
$D=i\sqrt{\Omega^{2}-C^{2}}$.

Hence (returning to the variable $x$)
\[
\mathcal{E}(x,t)=A\sqrt[4]{\mu}\sqrt{\alpha x+\beta}e^{i\Omega t}\left(
\left(  \frac{\alpha x+\beta}{\beta}\right)  ^{\frac{D\sqrt{\mu}}{\alpha}%
}+\frac{D+C}{D-C}\left(  \frac{\alpha x+\beta}{\beta}\right)  ^{-\frac
{D\sqrt{\mu}}{\alpha}}\right)
\]
and%
\[
\mathcal{H}(x,t)=\frac{A}{D-C}\frac{\Omega\,e^{i\Omega t}}{\sqrt[4]{\mu}%
\sqrt{\alpha x+\beta}}\left(  \left(  \frac{\alpha x+\beta}{\beta}\right)
^{\frac{D\sqrt{\mu}}{\alpha}}-\left(  \frac{\alpha x+\beta}{\beta}\right)
^{-\frac{D\sqrt{\mu}}{\alpha}}\right)
\]
satisfy the Maxwell system \eqref{Maxwell1+1} with the permittivity \eqref{ExEpsilon} and the initial conditions
\[
\mathcal{E}(0,t)=\frac{2AD}{D-C}\sqrt[4]{\mu}\sqrt{\beta}e^{i\Omega t}\quad\text{and}\quad
\mathcal{H}(0,t)=0.
\]

For the numerical calculation we considered an interval $[0,6]$ for both $x$
and $t$ and took $\alpha=2$, $\beta=1$, $\mu=1$. For the initial condition we took the sum of four terms, each of the form \eqref{Ex2Wxit} having $\Omega_1=-\Omega_2=C+1$, $\Omega_3=-\Omega_4=C+2$. Since the expression \eqref{Ex2Wxit} for $\xi=0$ reduces to $W(0,t)=\frac{2AD}{D-C}e^{i\Omega t}$, we took $A_i = \frac{D_i-C}{D_i}$, $i=1,\ldots,4$ and obtained the initial data $W^\pm_0(t)=4 \cos (C+1)t + 4\cos (C+2)t$.

We compared solutions computed by \eqref{complete sol} and \eqref{W_Nrearranged}. All calculations were performed using Matlab 2017 in the machine precision.
The exact expressions were used only for the function $\varepsilon(x)$ and its
derivatives, all other functions involved were computed numerically.

The permittivity $\varepsilon(x)$ was
approximated on uniform mesh of $5001$ points. The new variable $\xi$ was
obtained by the modified 6 point Newton-Cottes integration formula, see
\cite{KNT 2015} for details. The same integration formula was used for
calculation of the coefficients $a_n$ and $b_n$ and the integrals in \eqref{complete sol}. Note that the integration
with respect to the variable $\xi$ requires integration over a non-uniform
mesh, however such inconvenience can be easily avoided observing that
$\int_{0}^{\xi(x)} \widetilde g(\xi)\,d\xi= \int_{0}^{x} g(s) \xi^{\prime
}(s)\,ds = \int_{0}^{x} g(s)\sqrt{\mu\varepsilon(s)}\,ds$ for any function
$g(x)=\widetilde g(\xi(x))$.
To evaluate the indefinite integrals $\int z^k W_0^{\pm}(z)\,dz$, $k=0,\ldots,N$ we approximated the integrands as splines and used the function $\texttt{fnint}$ from Matlab. The main reason for such choice is that the set of values taken by $t-\xi$ and $t+\xi$ may be rather large and need not to be uniformly spaced and so some kind of interpolation is necessary, we opted for splines.

On Figure \ref{fig:Ex1} we show the permittivity (on the left) and the initial data $W_0^\pm(t)$  (on the right).

\begin{figure}[tbh]
\centering
\includegraphics[
width=3in,
height=2in]{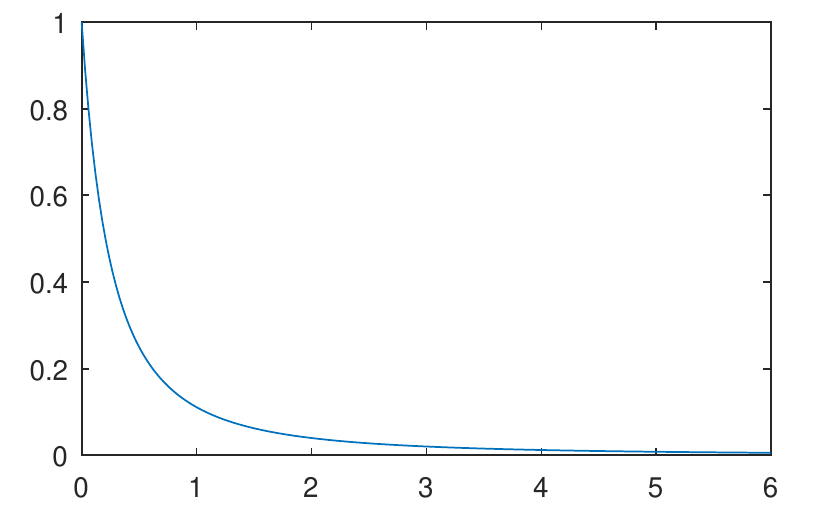} \quad\includegraphics[
width=3in,
height=2in]{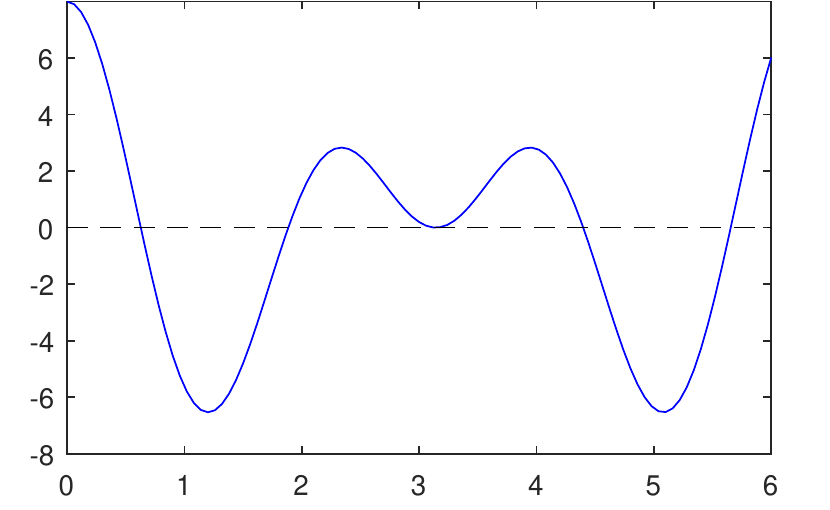}\caption{Graphs of the permittivity
$\varepsilon(x)$ (on the left) and the initial data $W^{\pm}_0(t)$ (on the right, real part in solid blue line, imaginary part in dashed black line) from
Example \ref{Ex1}.}
\label{fig:Ex1}%
\end{figure}

On Figure \ref{fig:Ex2} we show the graphs of the solutions $\mathcal{E}(x,t)$ (on the left) and $\mathcal{H}(x,t)$ (on the right, imaginary part). Note that for the chosen initial data the values of $\mathcal{E}(x,t)$ are real (so that $\mathcal{E}$ coincides with the component $E_2$), while the values of $\mathcal{H}(x,t)$ are purely imaginary (so that the imaginary part of $\mathcal{H}$ coincides with the component $H_3$)

\begin{figure}[tbh]
\centering
\includegraphics[
width=3in,
height=2.4in]{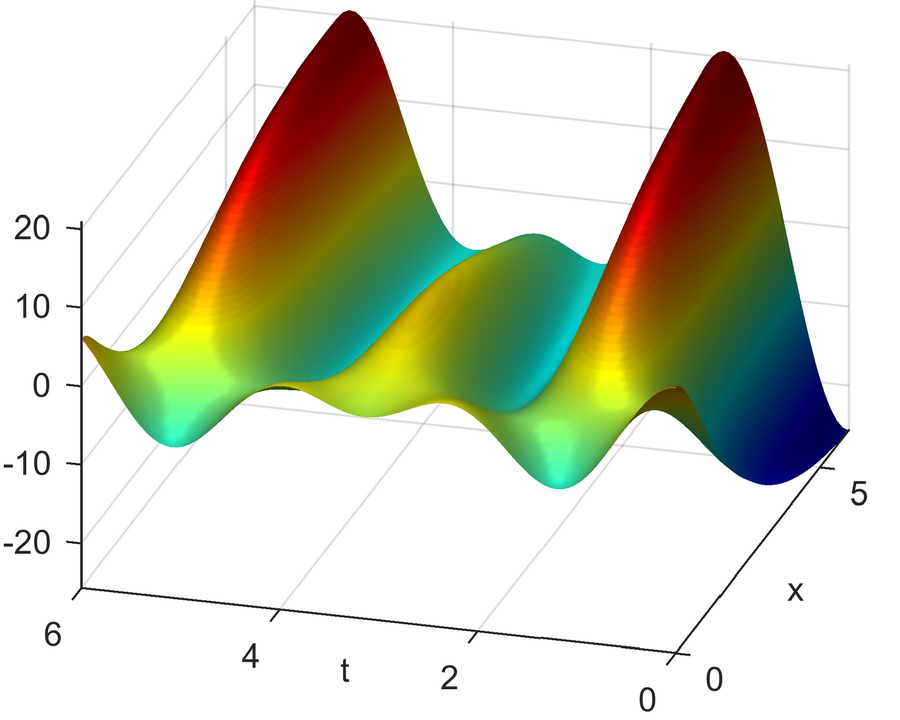} \quad\includegraphics[
width=3in,
height=2.4in]{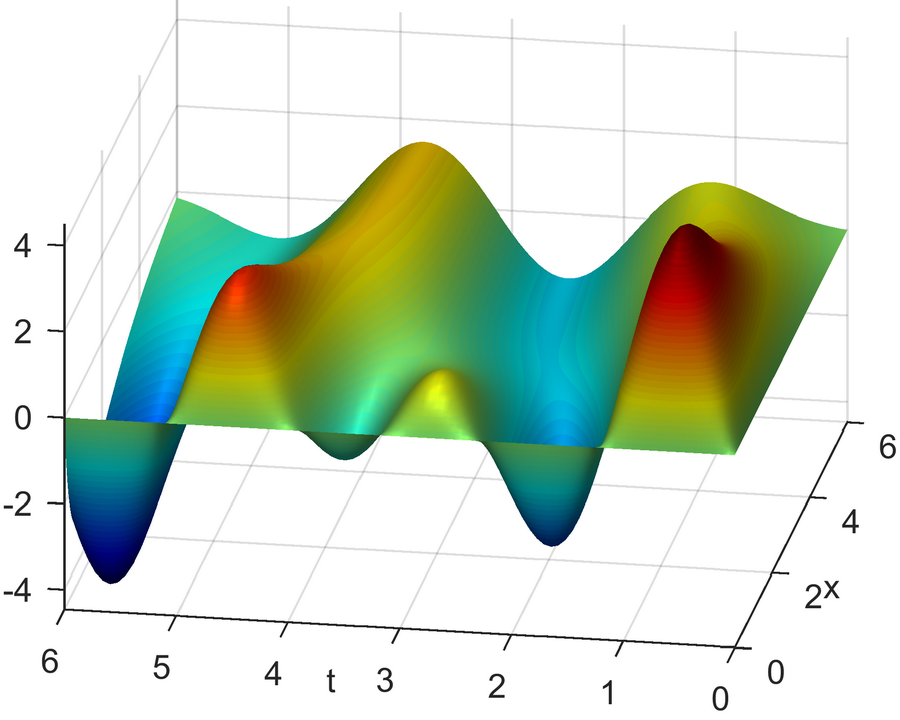}\caption{Graphs of the exact solutions
$\mathcal{E}(x,t)$ (on the left) and $\mathcal{H}(x,t)$ (on the right, imaginary part) from
Example \ref{Ex1}.}%
\label{fig:Ex2}%
\end{figure}

The developed program found the optimal value of $N$ for the approximation of the transmutation operators to be $N=13$ (see \cite{KNT 2015} for related details). On Figure \ref{fig:Ex3} we show the graphs of the absolute
errors of the computed $\mathcal{E}(x,t)$ and $\mathcal{H}(x,t)$ using directly formula \eqref{complete sol}. One can appreciate excellent accuracy, however the computation time of the solutions on the mesh of $201\times 101$ points $(x,t)$ was about 4 minutes.

\begin{figure}[tbh]
\centering
\includegraphics[
width=3in,
height=2.4in]{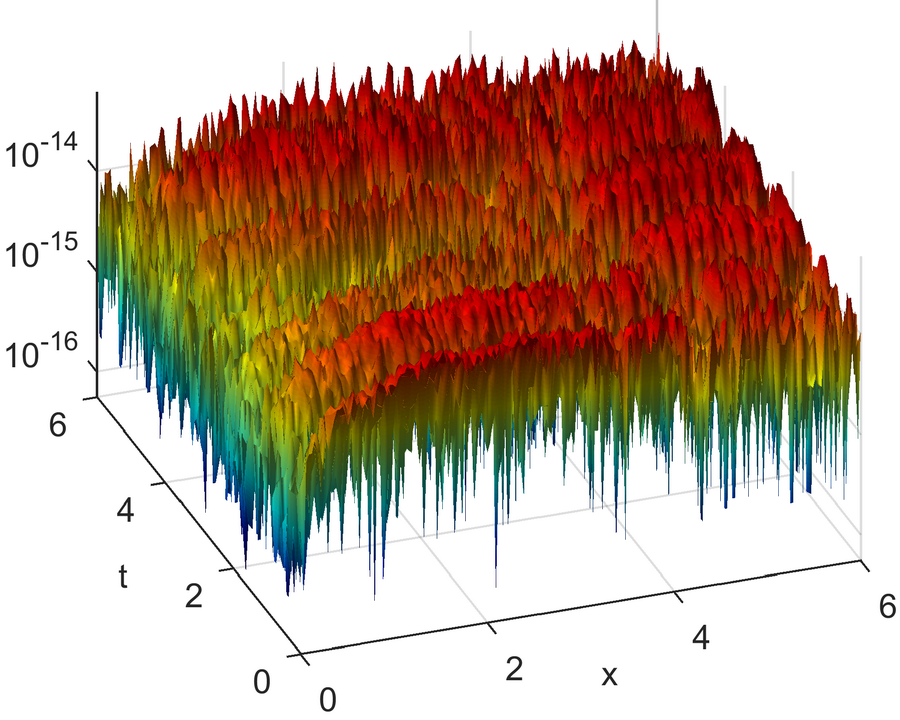} \quad\includegraphics[
width=3in,
height=2.4in]{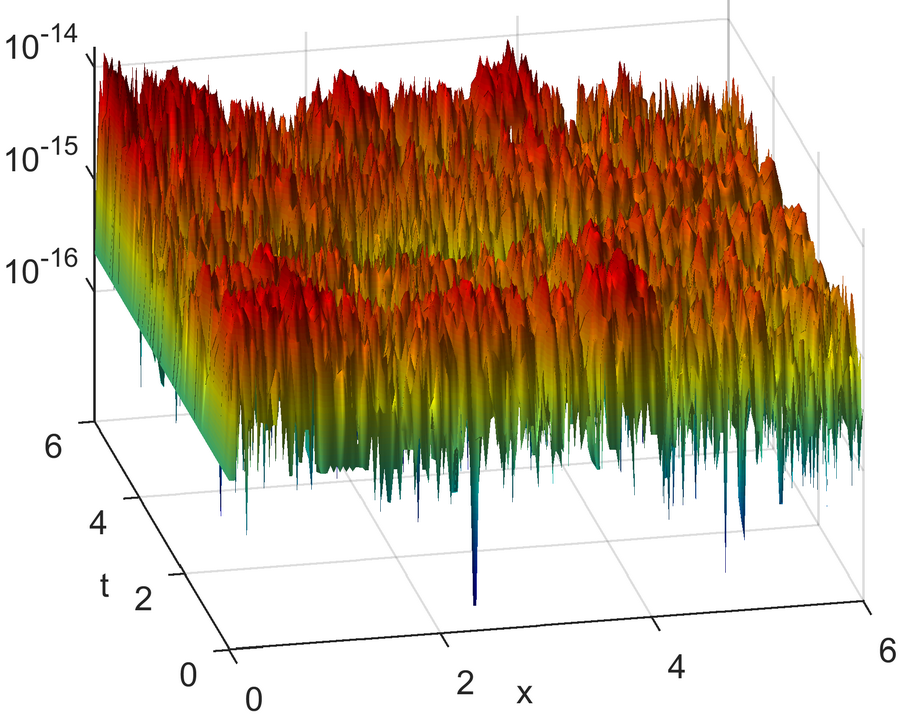}\caption{Graphs of the absolute errors of
$\mathcal{E}(x,t)$ (on the left) and $\mathcal{H}(x,t)$ (on the right) computed using formula \eqref{complete sol} truncated to $N=13$.}
\label{fig:Ex3}%
\end{figure}

The computation time required by formula \eqref{W_Nrearranged} was only 4 seconds, and most of this time was spent for the construction of splines. On Figure \ref{fig:Ex4} (top graphs) we show the graphs of the absolute errors of the computed $\mathcal{E}(x,t)$ and $\mathcal{H}(x,t)$. As one can see, the error is close to machine precision limit away from $x=0$, however rapidly increasing as $x$ approaches $0$ due to division by large powers of $\xi$ in the coefficients $c_{k,N}^\pm$.

The situation can be improved by taking smaller value of $N$ for this region. For example, for $N=6$ the maximum absolute error reduces to $7\cdot 10^{-8}$, however the errors for values of $x$ distant from 0 growth. See Figure \ref{fig:Ex4} (bottom row).

\begin{figure}[tbh]
\centering
$N=13$\\
\medskip
\includegraphics[
width=3in,
height=2.4in]{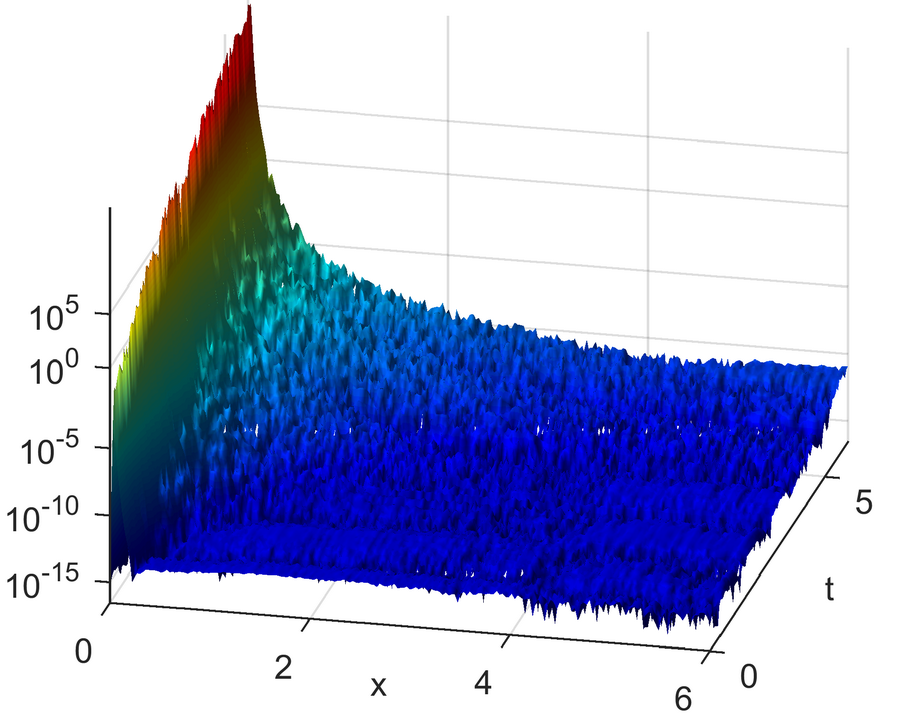} \quad\includegraphics[
width=3in,
height=2.4in]{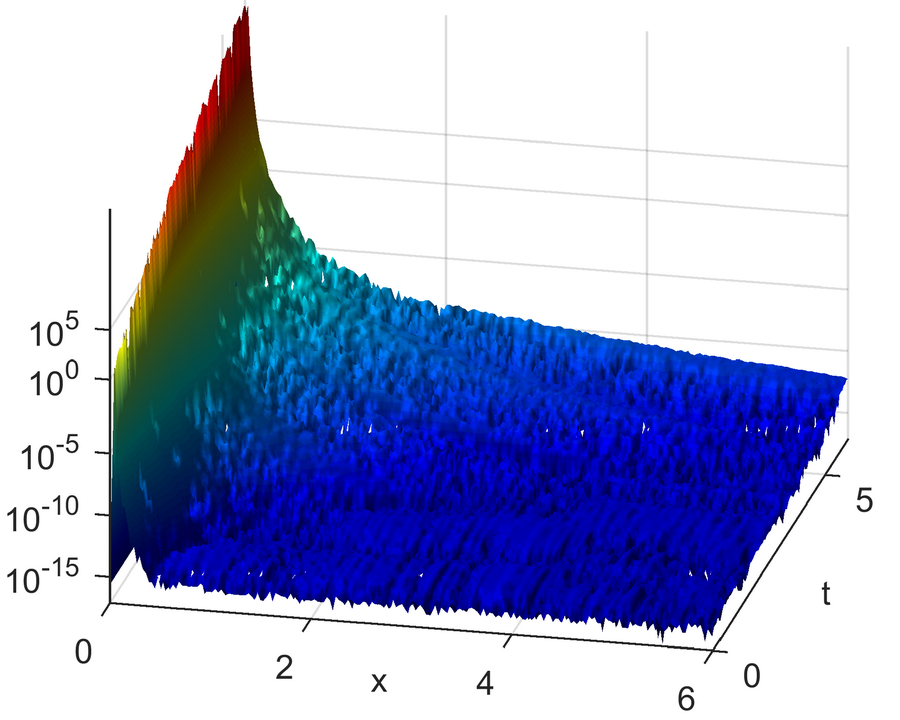}
\\
$N=6$\\
\medskip
\includegraphics[
width=3in,
height=2.4in]{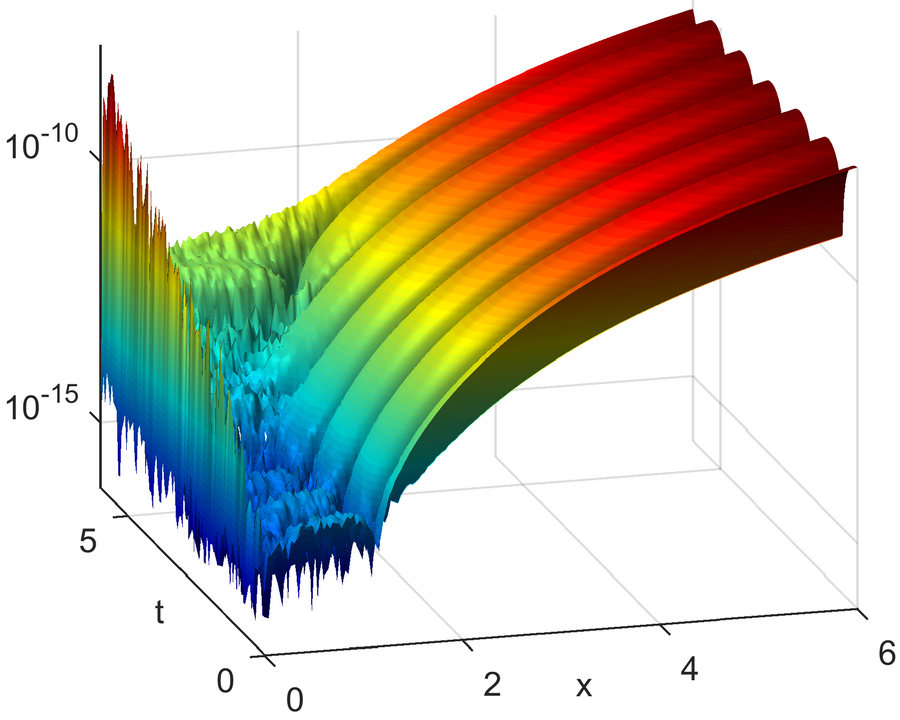} \quad\includegraphics[
width=3in,
height=2.4in]{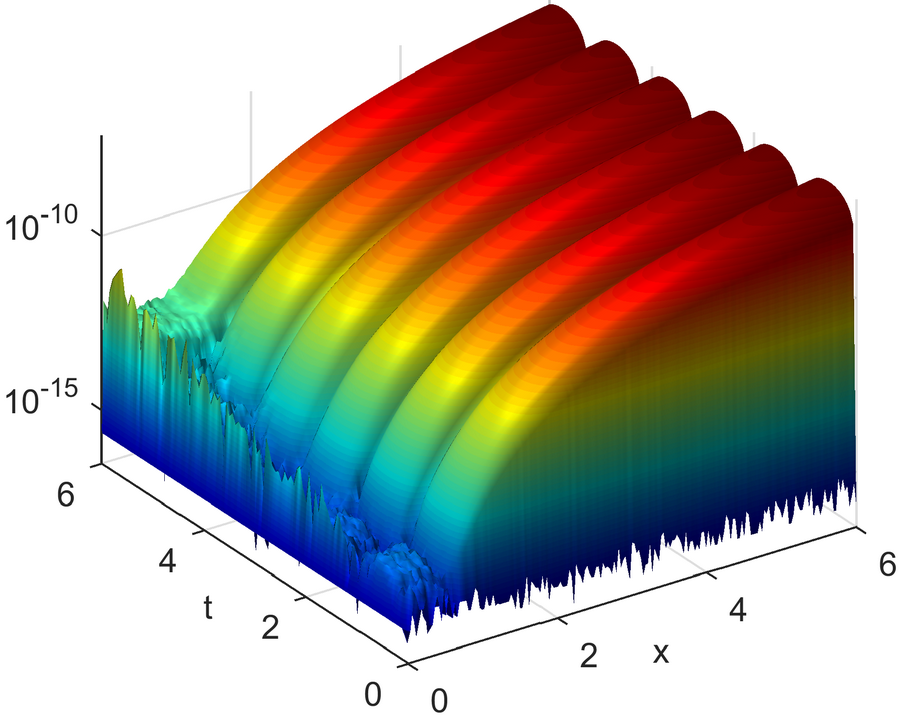}
\caption{Graphs of the absolute errors of
$\mathcal{E}(x,t)$ (on the left) and $\mathcal{H}(x,t)$ (on the right) computed using formula \eqref{W_Nrearranged}. Top graphs: using $N=13$, bottom graphs: using $N=6$.}
\label{fig:Ex4}
\end{figure}
\end{example}

\end{document}